\documentclass[prl,showpacs,twocolumn]{revtex4}
\usepackage{graphicx}
\usepackage{amsmath}
\usepackage{amssymb}
\usepackage{stmaryrd}
\usepackage{color}

\definecolor{labelkey}{cmyk}{.4,.2,0,0}

\newcommand{\be}{\begin{equation}}
\newcommand{\ee}{\end{equation}}
\newcommand{\bea}{\begin{eqnarray}}
\newcommand{\eea}{\end{eqnarray}}

\begin{document}

\title{Unbinding transition in semi-infinite two-dimensional localized systems}

\author{A. M. Somoza{$^1$}, P. Le Doussal{$^2$} and M. Ortu\~no{$^1$}}

\affiliation{{$^1$} Departamento de F\'{\i}sica-CIOyN,
Universidad de Murcia, Murcia 30.071, Spain}

\affiliation{{$^2$} CNRS-Laboratoire de
Physique Th{\'e}orique de l'Ecole Normale Sup{\'e}rieure, 24 rue
Lhomond,75231 Cedex 05, Paris France.}

\begin{abstract}
We consider a  two-dimensional strongly localized system defined in a half-space and whose transfer integral  in the edge can be different  than in the bulk. We predict an unbinding transition, as the edge transfer integral is varied, from a phase
where conduction paths are distributed across the bulk to a bound phase where propagation is mainly along the edge.
At criticality the logarithm of the conductance follows the $F_1$ Tracy-Widom distribution.
We verify numerically these predictions for both the Anderson and the Nguyen, Spivak and Shklovskii models.
We also check that for a  half-space, i.e., when the edge transfer integral is equal to the bulk transfer integral,  the distribution of the conductance is the $F_4$ Tracy-Widom distribution.
These findings are strong indications that random signs directed polymer models and their quantum extensions belong
to the Kardar-Parisi-Zhang universality class. 
We have analyzed finite-size corrections at criticality and for a half-plane.
\end{abstract}
\maketitle

The probability distribution function of the conductance $g$ of 
quantum random hopping models, such as the Anderson model,
has been much studied. In one dimension, it has been shown that
all the cumulants of $\ln g$ scale linearly with
system size \cite{Ro92}. Thus,
the distribution function of $\ln g$ approaches a Gaussian
form for asymptotically long systems and
is fully characterized by two parameters, the mean $\langle \ln g\rangle$
and the variance
$\sigma ^2 =\langle \ln^2 g\rangle- \langle\ln g\rangle^2$.
Both parameters are related to each other, supporting the extension of
the single parameter scaling  hypothesis \cite{AA79} to the distribution
function of the conductance \cite{shapiro}.

In two-dimensional (2D) systems, calculations of the conductance distribution function are possible
in the metallic regime, thanks to the non-linear sigma model \cite{2D}, but 
not until now in the strongly localized regime. 
However, two of us have argued, 
and demonstrated numerically, that in that regime $L \gg \xi$, $\xi$ being
the localization length, the
conductance takes the form \cite{SoOr07,PS05}
\begin{equation} 
\ln g=-\frac{2L}{\xi}+\alpha \left(\frac{L}{\xi}\right)^{1/3}\chi
\label{imp}\end{equation}
where  $\alpha$ is a constant and $\chi$ a random variable with a Tracy-Widom
(TW) cumulative distribution function (CDF). 
The TW distributions are  CDF for the largest eigenvalues
of large Gaussian random matrices \cite{TW}. It was found that the
random variable $\chi$ depends on the geometry of the model.
For narrow leads, the TW distribution associated to
the Gaussian unitary ensemble  $\chi=\chi_2$ is the CDF $F_2(x)$, i.e., 
${\rm Prob}(\chi_2 < x)=F_2(x)=F(x)^2$, 
where $F(x)$ is defined as $F(x)=\exp\{- \frac{1}{2} \int_x^{+\infty} (s-x) u(s)^2 ds \}$
and $u(x)$ is the solution  of the Painlev\'e II equation 
$u''(x)=2u(x)^3+xu(x)$, 
with $u(x)\rightarrow
- {\rm Ai}(x)$  as $x\rightarrow + \infty$, ${\rm Ai}(x)$ being the Airy function.

The chain of arguments leading to (\ref{imp}) goes as follows. 
It was argued by Nguyen {\it et al.} (NSS) \cite{NS85} that 
quantum interference effects in the localized regime are
faithfully described by retaining only the shortest or forward-scattering paths.
Medina and Kardar \cite{MK92} studied in detail the NSS model,
focusing on its formulation as a model of directed
polymers (DP) with non-positive Boltzmann weights. They 
found that the variance of the tunneling probability
increases with distance as $L^{2/3}$ for 2D systems.
This suggests that in 2D the DP with non-positive Boltzmann weights 
should be in the same universality class as the standard (i.e. positive weight) DP problem,
well known to exhibit a wandering exponent of $2/3$. A qualitative replica argument
led to the same conclusion \cite{MK92}.

In the mean-time, much progress happened in the study of growth models in the
 Kardar-Parisi-Zhang (KPZ)  universality class  \cite{KPZ}.
The standard DP problem belongs to this class (the
KPZ height $h$ maps onto $\ln Z$ where $Z$ is the
DP partition sum).
Remarkably, it was found that
the TW distribution arises at large time scales (i.e. long polymer length) in
discrete models in the KPZ class. For example, the height in the polynuclear growth model \cite{png},
 the optimal energy in DP models \cite{Johansson}, and the length of the longest increasing subsequence  in a random permutation \cite{BD99} all follow TW distributions.

This body of facts thus led to the conjecture (\ref{imp}) with the $1/3$ exponent
 of the KPZ class, and to its numerical verification.
More recently the KPZ equation and the DP problem have been solved directly in the
continuum, using integrability by the Bethe Ansatz of an associated quantum boson model 
\cite{we,dotsenko,spohnKPZEdge,corwinDP}.
Various boundary conditions where treated, suggesting new tests for the 
conjecture (\ref{imp}) with various lead geometries. In particular
the continuum DP problem in a  half-space was found \cite{we-halfspace} to lead to the $F_4$ TW distribution,
associated to the Gaussian simplectic ensemble,  in agreement with earlier
results for discrete models \cite{sasamotohalfspace,BaikSymPermutations}. 
This distribution verifies $F_4(x)=F(x) (E(x)^{-1} + E(x))/2$
where $E(x) = \exp\{\frac{1}{2} \int_x^{+\infty} u(s) ds\}$, $F(x)$ and $u(x)$ are the functions defined above. It was thus conjectured \cite{we-halfspace} 
that in a half-space the conductance near the edge should be of the form (\ref{imp})
with $\chi=\chi_4$  of CDF given by $F_4$.

The aim of this paper is to study  the logarithm of the conductance $\ln g$ in 2D systems in the strongly
localized regime in order to (i)  verify numerically the conjecture that the CDF
of for a half-plane is the $F_4(x)$ function, and (ii)
predict and analyze the transition from 2D to 1D 
as the hopping amplitude $t$ at the edge is increased, favoring propagation along the edge.
Exactly at criticality we observe that the CDF of $\ln g$ is the
TW distribution $F_1(x)=F(x) E(x)$ of the Gaussian orthogonal ensemble. 
This transition is a generalization of the unbinding transition studied for  positive weight DP
 \cite{Kardar,BaikSymPermutations}.
Remarkably, the full crossover distribution obtained there
fits our data for the whole parameter range,
providing a very delicate test that the random sign DP belongs
to the KPZ universality class. This is all the more precious since at present no integrable
system has been found to solve the NSS model.

\begin{figure}[htb]
\includegraphics[width=.23\textwidth]{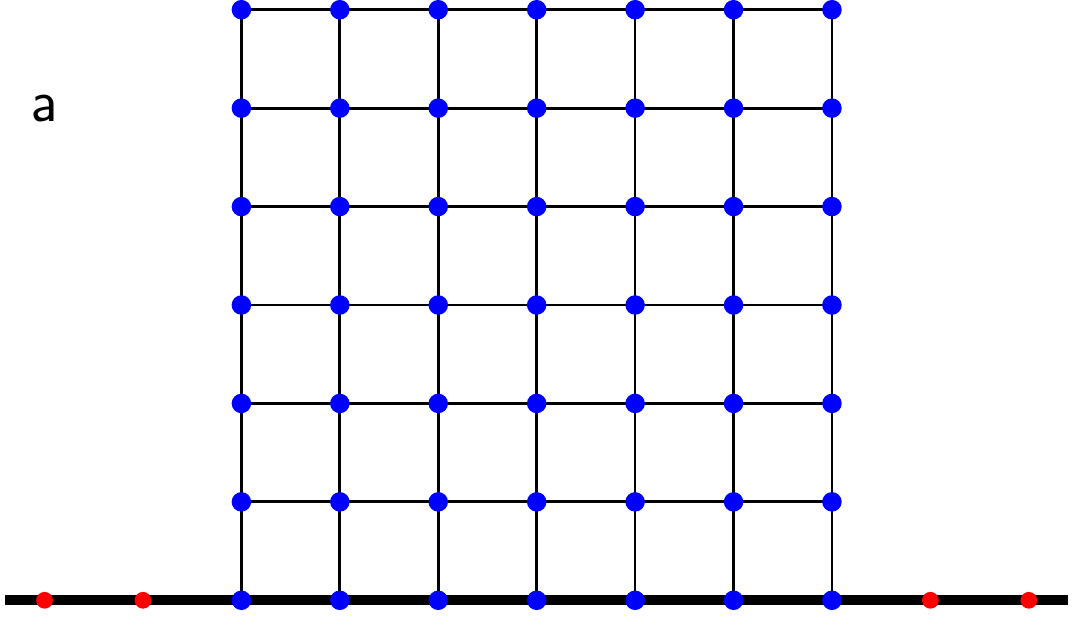}
\includegraphics[width=.23\textwidth]{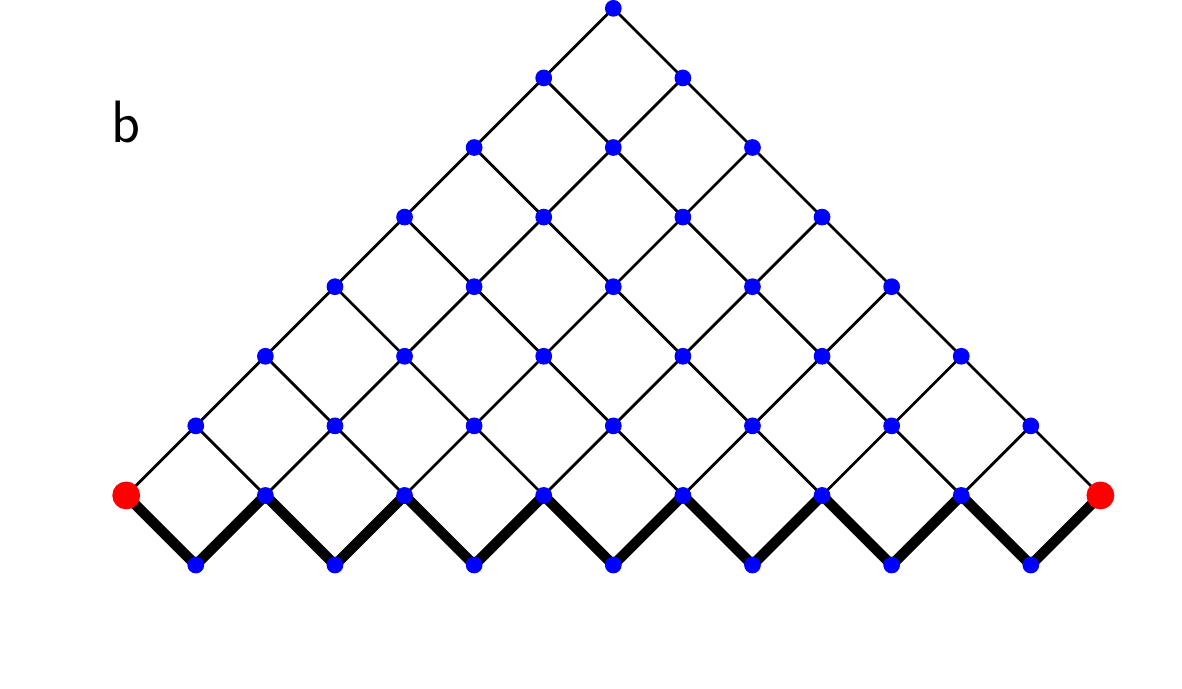}
\caption{Scheme of the geometrical arrangements of the a) Anderson model, and the b) NSS model.
Blue dots are site with random disorder energies  and red dots sites on the leads,
without disorder. Thin lines correspond to the bulk hopping strength equal 1, while thick
lines along the edge to $t$. }
\label{fig0}
\end{figure}

We focus on the Anderson model on a square sample of finite size
$L\times L$ described by the Hamiltonian
\begin{eqnarray}
H = \sum_{i} \epsilon_{i}a_{i}^{\dagger}a_{i}+\sum_{\langle i,j\rangle} t_{i,j}
a_{j}^{\dagger} a_{i}+ {\rm h.c.} \;, \label{hamil}
\end{eqnarray}
where the operator $a_{i}^{\dagger}$ ($a_{i}$) creates (destroys)
an electron at site $i$ of an square lattice and $\epsilon_{i}$ is
the energy of this site chosen at random $\epsilon_{i}\in [-W/2,W/2]$, with
$W$  the strength of the disorder. The double sum runs over nearest neighbors. The
hopping matrix element $t_{i,j}$ is taken equal to $1$ everywhere, except
between the sites along one edge where is equal $t>1$ (see figure \ref{fig0}).
The value 
in the bulk sets the energy scale, while the lattice constant sets the
length scale. The unit of conductance is $2 e^2/h$.

For the Anderson model we have calculated the conductance through
Landauer's formula in terms of the transmission between perfect
leads, arranged as schematically shown in Fig.\ \ref{fig0}a. This is obtained from the Green function, which can be
calculated propagating layer by layer 
\cite{M85}. It is convenient to propagate along
the direction perpendicular to the leads, starting from the opposite edge,
so that each calculation of the bulk Green function can be used for any
value of  $t$.

The Green function between two sites
$a$ and $b$ can be written in terms of the locator expansion
\begin{equation}
\langle a|G|b\rangle =\sum_{\Gamma}\prod_{i\in\Gamma}\frac{1}{E-\epsilon_i} \;,
\label{mk2}
\end{equation}
where the sum runs over all possible paths connecting the two sites $a$ and $b$.
The NSS model assumes that, in the strongly localized regime,
(\ref{mk2}) is dominated by the forward-scattering paths and only takes into account
the contributions of such paths to the transmission amplitude between opposite corners of a square lattice.
As we want to study a half-space,
we consider a triangular sample as represented in Fig.\ \ref{fig0}b.
In order to mimic  the Anderson model, we choose the site disorder energy at random in the interval $\epsilon_{i}\in [-W/2,W/2]$, but if $|E-\epsilon_{i}|<1$,  
we set $1/|E-\epsilon_{i}|=1$, 
 partially incorporating  multiple scattering effects.
We note that at least for the planar symmetry this choice of disorder does no change the universality of the  problem \cite{epl}.

The calculation of the quantum amplitude in the NSS model is formally similar to the calculation of the partition
function of DP in a random potential
\begin{equation}
Z =\sum_{\Gamma}\exp\left\{-\beta\sum_{i\in\Gamma}h_i\right\} \;,
\label{pf}
\end{equation}
where $\beta=1/kT$, $h_i$ is a random site energy and $\Gamma$ runs over all possible
configurations of the DP. Equations (\ref{mk2}) and (\ref{pf}) are
equivalent provided that we can identify $-\beta h_i$ with $\ln (E-\epsilon_i)$.
While in DP the disorder energies $h_i$ are real, in the quantum case
$E-\epsilon_i$ does not have to be positive, implying a more general problem.

\begin{figure}[htb]
\includegraphics[width=.43\textwidth]{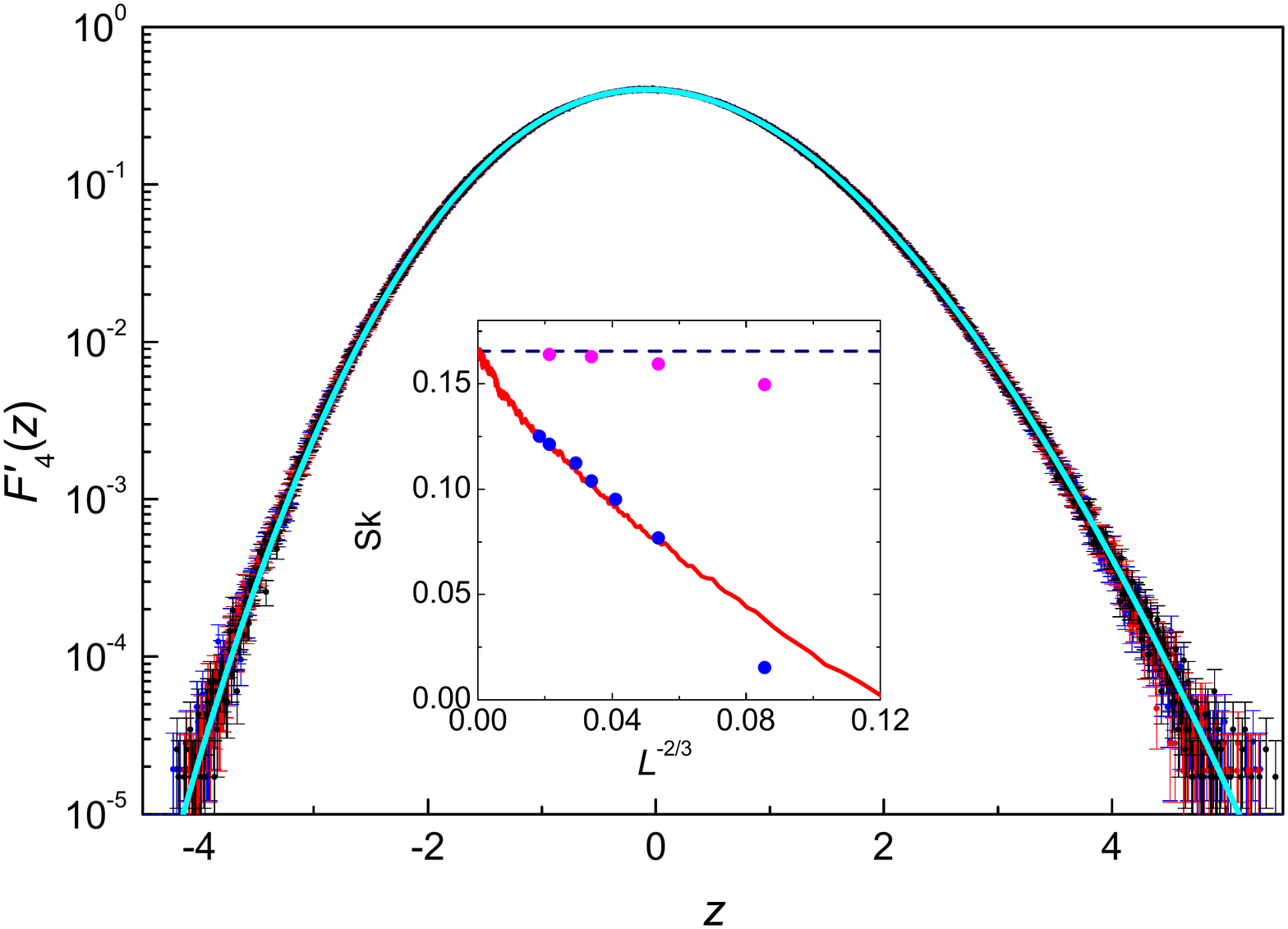}
\caption{ Histogram of $\ln g$ versus the scaled variable $z$
for three sizes and a disorder $W=10$ of the NSS model. 
The continuous line corresponds to $F_4'(z)$. Inset: skewness
vs.  $L^{-2/3}$ for the NSS model (red curve) for the Anderson model with $W=10$ (blue dots) and 20 (magenta dots);  the dashed line is Sk$_4$. } \label{fig1}
\end{figure}

It is interesting to study first the conductance distribution of  the NSS and Anderson models in a half-space and check whether the conjecture of Ref. \cite{we-halfspace}, based on the exact result for the continuum DP, that the CDF of $\chi$ is the TW function $F_4$ is verified by our two models.
In Fig.\ \ref{fig1} we plot  histograms of $\ln g$ for the NSS model as a function
of $z=(\ln g-A)/B$, where $A$ and $B$ are chosen so that $z$ has the same mean and variance
as $F_4$. 
The lateral dimensions are $L=2500$ (blue dots), 5000 (red dots) and 10000 (black dots), and
the number of realizations is  $4 \times 10^6$. 
The solid line  corresponds to the TW function $F_4'$.
We see a perfect agreement between our numerical results and the TW distribution
for more than four orders of magnitude.
A  similar agreement is found for the Anderson model (not shown), taking into account that it corresponds to smaller system sizes.

The inset of Fig.\ \ref{fig1} we plot the skewness of  $\ln g$ versus  $L^{-2/3}$, which according to the discussion below is expected to be the leading order correction. 
The red line corresponds to the NSS model, while the dots to the Anderson model ($W=10$ blue, 20 magenta).
The dashed line is   Sk$_4=0.165509...$ The skewness of the numerical results for both models tends to the predicted value.

We now argue theoretically that the NSS and the Anderson models exhibit a phase transition
between (i) a 2D phase 
where conduction paths wander unboundedly in the
half-space and the fluctuations of $\ln g$ at large $L$ are described by (\ref{imp}) with $\chi=\chi_4$,
and (ii) a quasi-1D localized
phase where the conduction paths have 1D character at large scale, and $\ln g$ has a log-normal
distribution with fluctuations scaling as $L^{1/2}$,
 i.e.  (\ref{imp}) still holds
but with a size dependent variable $\chi \to \chi_L$, scaling now as $\chi_L \sim L^{-1/6}$. 
Exactly at the transition $t=t_{\rm c}$, the fluctuations of $\ln g$ at large $L$ are described by (\ref{imp}) with now $\chi=\chi_1$
distributed according to $F_1$. Hence the 1/3 KPZ exponent holds up to and at the transition.
This transition can be described as an {\it unbinding transition} of conduction paths.
In  the standard DP the unbinding transition was predicted in \cite{Kardar} and worked 
out in detail in the framework of symmetrized random permutations in \cite{BaikSymPermutations}.
Before using these results below, let us first give the physical picture and the scaling arguments.

In the {\it unbound phase}, $t>t_{\rm c}$,  the paths wander in the 2D space, but come back from
time to time to the boundary (they still feel the boundary hence the change from $F_1$ to $F_4$). 
Their typical transverse wandering is $L^{2/3}$. 
Ins the {\it bound phase}, the paths return to the wall, 
with a typical length  $\xi_u$ which diverges at the transition as $\xi_u \sim (t-t_{\rm c})^{-3}$.
Near the transition there are thus $L/\xi_u$ independent pieces of paths, each fluctuating as $\xi_u^{1/3}$
hence the variance should behave as $\overline{ (\ln g)^2 }^c = \xi_u^{2/3} (L/\xi_u) = L/\xi_u^{1/3}\sim (t-t_{\rm c}) L$.
Far from the transition this picture predicts that the skewness
should decay as Sk$\sim (L/\xi_u)^{-1/2}$, since in the 1D phase the central limit theorem holds for $\ln g$ and 
all cumulants scale as $L$ i.e. $\overline{(\ln g)^n}^c \sim L$.

Let us now consider the critical scaling around the transition point \cite{BaikSymPermutations}.
Our prediction is that $\ln g$ always behaves as in (\ref{imp}) with the random variable $\chi$  having a CDF
Prob$(\chi<x)={\cal F}(x,w=- c \tilde w)$ that depends on the scaling variable 
$\tilde w=  (t-t_{\rm c}) L^{1/3}$, $c>0$ being an (unknown) proportionality constant. 
The universal
crossover function ${\cal F}(x,w)$ (corresponding to $F^{\boxslash}(x,w/2)$ in Ref. \onlinecite{BaikSymPermutations}) satisfies 
\begin{eqnarray}
&& {\cal F}(x,w) = \frac{1}{2} F(x) \big( (f(x,w) - g(x,w)) E(x)^{-1} \label{sca} \\
&& + (f(x,w) + g(x,w)) E(x) \big) \nonumber  
\end{eqnarray}
where the functions $f(x,w)$ and $g(x,w)$ are the ones of \cite{Baik2} and verify:
\begin{eqnarray}
 \partial_w f(x,w)& =& u(x)^2 f(x,w) - (w u(x) + u'(x)) g(x,w) \nonumber 
 \\
 \partial_w g(x,w) &=& - (w  u(x) - u'(x)) f(x,w) +\nonumber\\
&& (w^2-x-u^2(x)) g(x,w) \label{sca2}
\end{eqnarray}
and are subjected to the initial condition at criticality 
$f(x,0)=-g(x,0)=E(x)^2$. 
Then, the CDF at criticality is ${\cal F}(x,0)=F_1(x)$.
In the limit $w\to +\infty$, one has    
$f \to 1$ and $g \to 0$, hence ${\cal F}(x,\infty)=F_4$. In the opposite limit,
 $w \to -\infty$ \cite{Baik2}
${\cal F}(w^2 + y \sqrt{|w|} , w) \to {\rm erf}(y)$ 
and one recovers Gaussian fluctuations for $\ln g$.

\begin{figure}[htb]
\includegraphics[width=.43\textwidth]{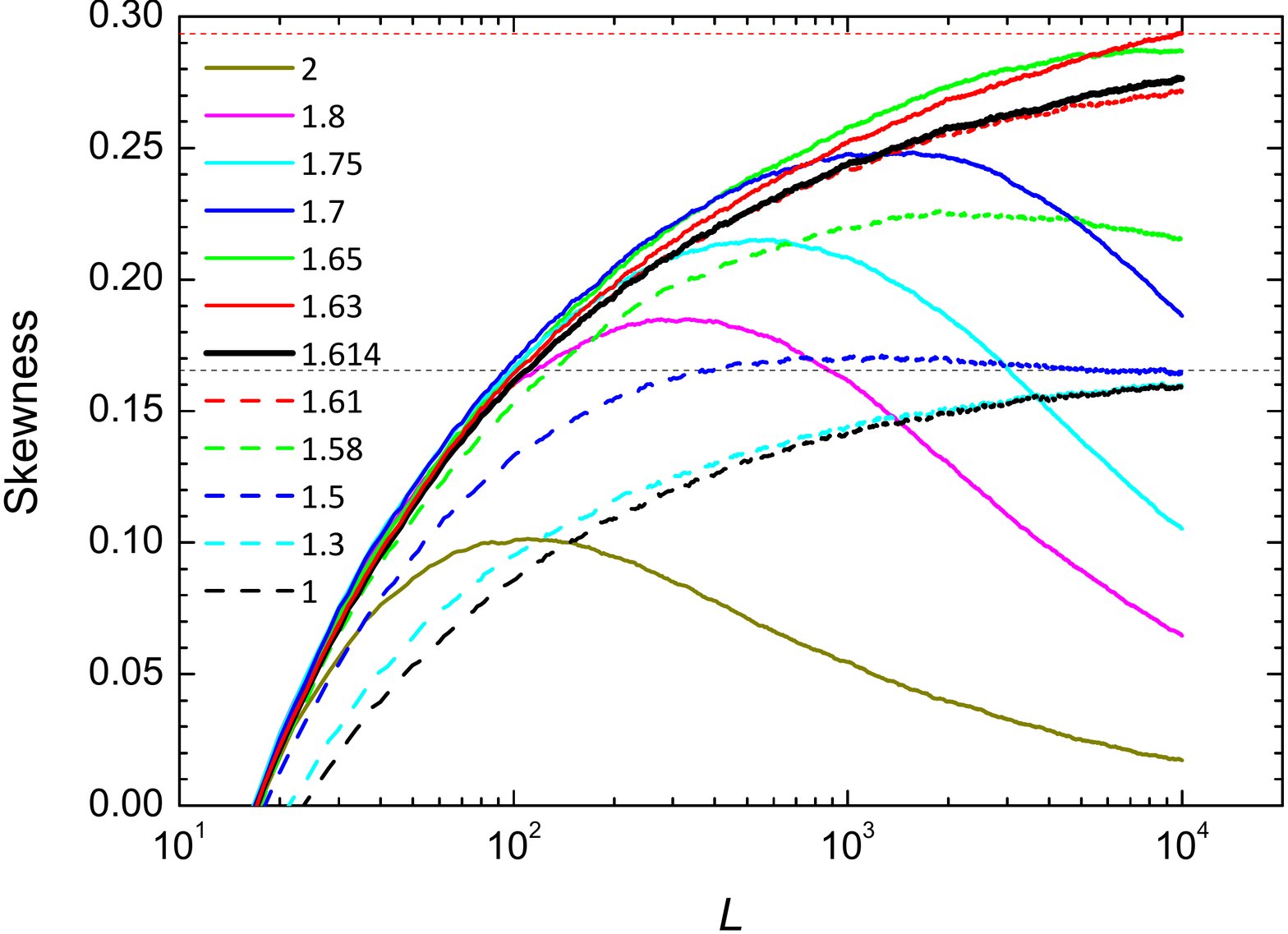}
\caption{Skewness versus $L$ on a logarithmic scale for the values of $t$ shown
in the figure. The red dashed line corresponds to Sk$_1$ and the dashed black horizontal line to Sk$_4$.} \label{fig2}
\end{figure}

To study the transition and to characterize the two phases, we  analyze the skewness of the distribution of $\ln g$ for the NSS model as we vary  $t$. 
The skewness should tend
to 0 in the localized phase, to Sk$_1=0.293464...$ at the transition and to Sk$_4$
in the unbound phase.
In Fig.\ \ref{fig2} we plot the skewness as a function of size on a logarithmic scale for
several values of the edge strength $t$. The upper horizontal line corresponds to the expected value
of the skewness at the critical point, Sk$_1$, and the lower horizontal line to the limiting value in the bulk phase,  Sk$_4$.  The black curve represents  the skewness at the critical value  $t_{\rm c}\approx 1.613(1)$ and tends to the upper horizontal  line,  Sk$_1$. Solid curves are for $t>t_{\rm c}$ and after reaching a maximum start decreasing, eventually tending to zero.
Dashed lines correspond to $t<t_{\rm c}$  and tend to Sk$_4$, middle horizontal line in Fig.\ \ref{fig2}.

\begin{figure}[htb]
\includegraphics[width=.45\textwidth]{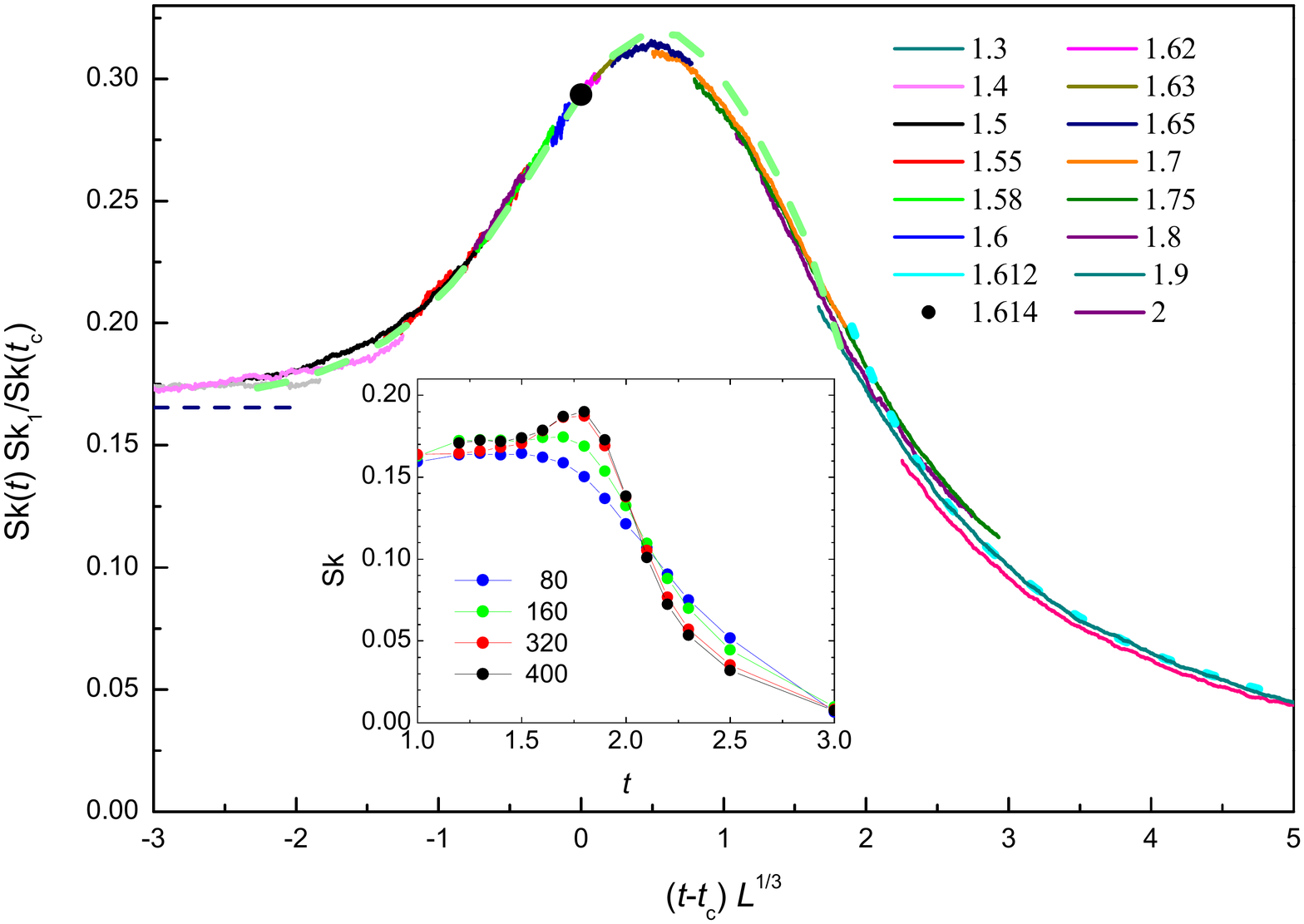}
\caption{Skewness scaled as a function of ${ \tilde w}=(t-t_{\rm c})L^{1/3}$ for several values of  $t$. The black dot corresponds to Sk$_1$ at the transition  and the horizontal dashed line to Sk$_4$. The dashed curve is the theoretical prediction for the scaling function and the dotted curve the predicted asymptotic behavior at large $\tilde w$.
Inset: skewness as a function of $t$ for the Anderson model.} \label{fig3}
\end{figure}

It is possible to scale the raw data for the skewness at large values of $L$ into a universal curve using as scaling variable ${\tilde w}=(t-t_{\rm c})L^{1/3}$. Looking at the behavior of the skewness for the critical value $t_{\rm c}$ in Fig.\ \ref{fig2}, it is clear that finite size effects are important. 
To take them into account we   assume Sk$(t,L) \approx \mbox{Sk}(\tilde w) f(L)$, where $f(L)$ incorporates finite size effects in a simple way   $f(L) = \mbox{Sk}(t_{\rm c},L)/\mbox{Sk}_1$.
In Fig.\ \ref{fig3} we plot this renormalized values Sk$(\tilde{w})$ 
for several values of $t$. All curves are plotted for the range $200<L<10^{4}$. The black dot represents the transition point and so is equal Sk$_1$ and is placed at the vertical axis, $t=t_{\rm c}$. The dashed line is the theoretical prediction for the scaling function, i.e., the solution of Eqs.\  
(\ref{sca2}), using a value $c=0.9$ for the multiplicative constant. 
It presents a maximum in the localized phase  well reproduced by our simulations.
The blue dotted line on the right represents the limiting behavior in the bound phase, $b\tilde w^{-3/2}$, with the fitting parameter $b=0.55$.
The overlap and the agreement with theoretical expectations is quite good,
specially noting that the only free parameter to overlap all the curves is  $t_{\rm c}$.

The results for the Anderson model confirm all predictions, although  the maximum size that can be calculated, $L=400$, is still relatively far from the asymptotic behavior.
In the inset of Fig.\ \ref{fig3}, we plot the skewness as a function of $t$ for several system sizes of the Anderson model and a disorder $W=20$.
A peak around  $t\approx 1.8$ develops with size, but its maximum is still far from Sk$_1$.  This results are fully consistent with NSS for similar sizes.

\begin{figure}[htb]
\includegraphics[width=.4\textwidth]{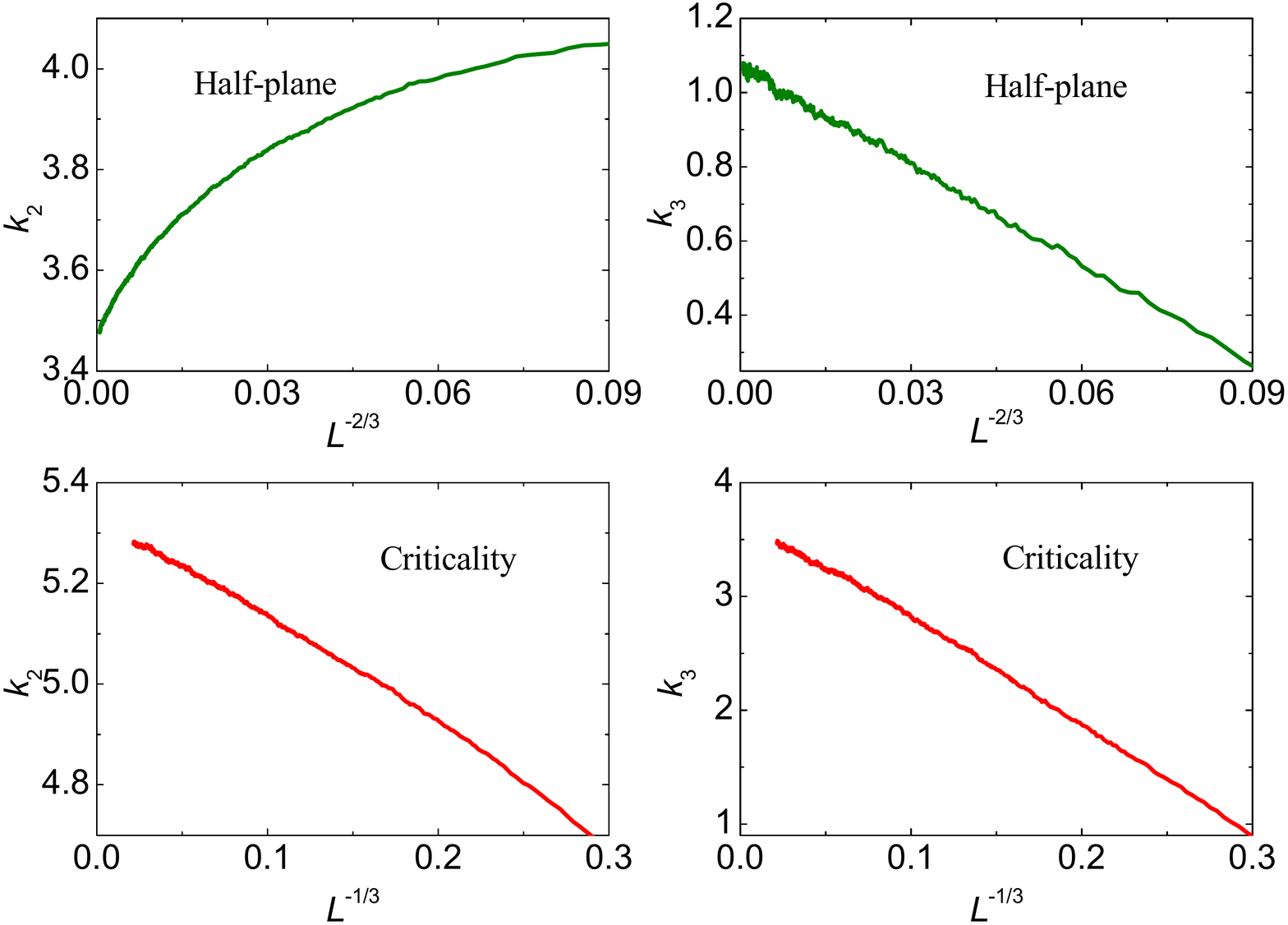}
	\caption{Leading order corrections of the second $k_2$ (left) and third $k_3$ (right) cumulants in the NSS model for a half-space versus $L^{-2/3}$ (top panels) and at the unbinding transition versus  $L^{-1/3}$ (bottom panels). 
\label{new}}
\end{figure}

It is interesting to compare the finite size corrections in our models with those for discrete growth models and experiments in the KPZ class.
Consider $K_n$ the cumulant of $\ln g$ (here) and of the interface height $h$ (there, time being denoted as $L$) and
the scaled cumulants $k_n=K_n/L^{n/3}$ which converge to constants.
The finite size corrections of the $k_n$ where found (there in the {\it bulk}) to scale
as $L^{-1/3}$ for $n=1$ and as {\it at most} $L^{-2/3}$ for all $n \geq 2$ \cite{Takeuchi,Ferrari}.
We study these corrections for the NSS model.
In the top panels of Fig. \ref{new} we plot $k_2$ (left) and $k_3$ (right) versus $L^{-2/3}$ for  the half-space, $t=1$.
We note that $k_3$ follows a very good linear behavior, while $k_2$ shows some curvature.
Based on the curvature of $k_2$ it is difficult to determine its leading finite size corrections exponent,
but the exponent $-2/3$ produces an extrapolated value for the skewness (Sk$=0.166$) very close the the theoretical expectation, 
indicating that finite size corrections for the half-space are similar to those observed for bulk growth models in KPZ class. 
In the lower panels of Fig.\ \ref{new} we plot $k_2$ (left) and $k_3$ (right) at the critical point versus $L^{-1/3}$.
The smooth behavior of both curves indicate that finite size corrections exhibit a distinct and novel behavior at criticality.
From the intercepts with the vertical axis of linear fits at large $L$, we obtain Sk$=0.297$, confirming that the CDF is the TW
function $F_1$.
The histograms of  $\ln g$ 
 at the critical point agree fairly well with these theoretical predictions.

We have shown that the NSS and the Anderson models in a half-plane
undergo an unbinding transition as the hopping amplitude at the edge is varied.  We show evidence that the analytical expressions for the conductance distribution functions and the scaling functions at the unbinding transitions in localized two-dimensional quantum systems are obtained from Eqs. (\ref{sca}-\ref{sca2}). 
We have also studied the  corresponding finite-size corrections. 
Similar unbinding transitions can occur in 3D systems when conduction through a surface or a line is favored. They may be relevant to the behavior of edge states.

\acknowledgements
We thank J. Baik   and K. A. Takeuchi for a useful exchange. 
AS and MO acknowledge financial support from the Spanish DGI and FEDER Grant No. FIS2012-3820.
PLD acknowledge support from PSL grant ANR-10-IDEX-0001-02-PSL.
PLD and MO thank KITP for hospitality and partial 
support under grant NSF PHY11-25915.

\end{document}